\documentclass[twocolumn,twocolappendix]{aastex631}

\def\degree{${}^{\circ}$}

\let\oldAA\AA
\renewcommand{\AA}{\text{\normalfont\oldAA}}
\usepackage{slantsc}
\usepackage{ulem}
\usepackage{scalerel}

\usepackage{float}
\usepackage{booktabs} 
\usepackage[flushleft]{threeparttable}
\usepackage{CJK}
\usepackage{CJK}

\begin{document}
\begin{CJK*}{UTF8}{gbsn}

\title{Abundant Molecular Gas in the Central Region of Lenticular Galaxy PGC 39535}

\author[0009-0002-5336-5962]{Jiantong Cui (崔健童)}
\affiliation{School of Astronomy and Space Science, Nanjing University, Nanjing 210093, People's Republic of China}
\affiliation{Key Laboratory of Modern Astronomy and Astrophysics (Nanjing), Ministry of Education, Nanjing 210093, People's Republic of China}

\author[0000-0002-3890-3729]{Qiusheng Gu}
\affiliation{School of Astronomy and Space Science, Nanjing University, Nanjing 210093, People's Republic of China}
\affiliation{Key Laboratory of Modern Astronomy and Astrophysics (Nanjing), Ministry of Education, Nanjing 210093, People's Republic of China}


\author[0000-0001-5988-2202]{Shiying Lu}
\affiliation{School of Astronomy and Space Science, Nanjing University, Nanjing 210093, People's Republic of China}
\affiliation{Key Laboratory of Modern Astronomy and Astrophysics (Nanjing), Ministry of Education, Nanjing 210093, People's Republic of China}

\author[0000-0003-1583-7404]{Zhengyi Chen}
\affiliation{School of Astronomy and Space Science, Nanjing University, Nanjing 210093, People's Republic of China}
\affiliation{Key Laboratory of Modern Astronomy and Astrophysics (Nanjing), Ministry of Education, Nanjing 210093, People's Republic of China}

\author[0000-0002-8437-6659]{Can Xu}
\affiliation{School of Astronomy and Space Science, Nanjing University, Nanjing 210093, People's Republic of China}
\affiliation{Key Laboratory of Modern Astronomy and Astrophysics (Nanjing), Ministry of Education, Nanjing 210093, People's Republic of China}

\author[0000-0002-0182-1973]{Zeyu Gao}
\affiliation{Department of Astronomy, School of Physics, Peking University, 5 Yiheyuan Road, Beijing 100871, People’s Republic of China}
\affiliation{Kavli Institute for Astronomy and Astrophysics, Peking University, Beijing 100871, People's Republic of China}

\correspondingauthor{Qiusheng Gu}
\email{qsgu@nju.edu.cn}


\begin{abstract}

Lenticular galaxies (S0s) in the local universe are generally absent of recent star formation and lack molecular gas. In this paper, we investigate one massive ($M_*$$\sim$5$\times10^{10}$ M$_\odot$) star-forming S0, PGC 39535, with the Northern Extended Millimeter Array (NOEMA). Using optical data from SDSS-IV MaNGA survey, we find star formation mainly concentrates in the central region of PGC 39535. The total star formation rate estimated using extinction-corrected H$\alpha$ flux is 1.57 M$_\odot$ yr$^{-1}$. Results of NOEMA observation suggest that the molecular gas mainly concentrates in the central regions as a gaseous bar and a ring-like structure, and shows similar kinematics as the stellar and ionized gas components. The total molecular gas mass estimated from CO(1-0) is (5.42$\pm$1.52)$\times$10$^{9}$ M$_{\odot}$.
We find PGC 39535 lies on the star-forming main sequence, but falls below Kennicutt-Schmidt relation of spiral galaxies, suggesting that the star formation efficiency may be suppressed by the massive bulge. The existence of a second Gaussian component in the CO spectrum of the central region indicates possible gas flows. Furthermore, our analyses suggest that PGC 39535 resides in the center of a massive group and the derived star formation history indicates it may experience a series of gas-rich mergers over the past 2$\sim$7 Gyr.  

\end{abstract}



\section{Introduction} \label{sec:intro}

Lenticular galaxies (S0s) contain both a bulge and a disk, lacking prominent spiral structures and serving as a bridge between elliptical and spiral galaxies in the well-known `tuning fork' diagram \citep{Hubble_1936}. In most cases, the bulges of S0s are red and lack cold gas \citep[e.g.,][]{Bregman_1992,Sandage_1978}. Generally, S0s exhibit lower specific star formation rates than main sequence galaxies\citep[e.g.,][]{Caldwell_1993}.

S0s are usually suggested to originate from two pathways \citep[e.g.,][]{Deeley_2020,Deeley_2021}. In dense environments, such as clusters and groups, they are believed to form from faded spirals \citep[e.g.,][]{D'Onofrio_2015}. Throughout this process, ram pressure stripping plays a key role to remove the gas in the disk through interaction with the intergalactic medium \citep[e.g.,][]{Gunn_1972,Quilis_2000}. Other processes also can have an impact on the gas removal, including galaxy harassment \citep[e.g.][]{Moore_1996}, gravitational tidal effects and mergers \citep[e.g.,][]{Barnes_1992,Mazzei_2014a,Mazzei_2014b}. Observations of spirals transitioning into S0s provide direct evidence for this scenario \citep[e.g.,][]{DeGraaff_2007,Loni_2023}. The fraction of S0s in clusters is found to decrease with redshift, which anticorrelates with that of spiral galaxies \citep[e.g.,][]{Dressler_1980,Dressler_1997,Postman_2005}. Since gas stripping often occurs at the outskirts of galaxy disk, S0s formed in high density environments often exhibit blue cores and rotationally supported dynamics \citep[e.g.,][]{Deeley_2020,Deeley_2021}.
S0s in the field are supposed to have different properties from those in high density environment. They tend to have low v/$\sigma$ and misaligned stellar and gas components. These S0s are thought to form from merger event \citep[e.g.,][]{Bournaud_2005,Bassett_2017,Eliche-Moral_2018}.

Although S0s show old stellar populations, it is found that there exists gas in a large number of S0s \citep[e.g.,][]{Welch_2003,Welch_2010}, supporting recent star-forming activities \citep[e.g.,][]{Xiao_2016,Sil'chenko_2019}. 
Star-forming S0s in different environments are affected by different factors.
In the groups, star-forming S0s are thought to be successors of faded spirals while in the field the star formation may result from gas accretion or minor mergers \citep[e.g.,][]{Xu_2022}.

Additionally, the stellar mass might be a key factor on the star formation as well. 
The observational evidence indicates a `downsizing' trend in galaxy formation \citep[e.g.][]{Cowie_1996,Heavens_2004,Juneau_2005,Thomas_2010,Perez-Gonzalez_2008}, where stars in massive galaxies are observed to form early within a short timescale. One possible explanation \citep{Cattaneo_2008} for this phenomenon is that galaxies may be unable to accrete or retain cold gas in massive halos, either due to AGN feedback \citep[e.g.,][]{Sanchez_2018,Belfiore_2018} or viral heating \citep[e.g.,][]{Dekel_2006}. Consequently, massive galaxies in the local universe typically lack ongoing star-forming activities. Therefore, studying the properties of star-forming massive galaxies can help us understand galaxy formation and evolution better.

\begin{figure}
\includegraphics{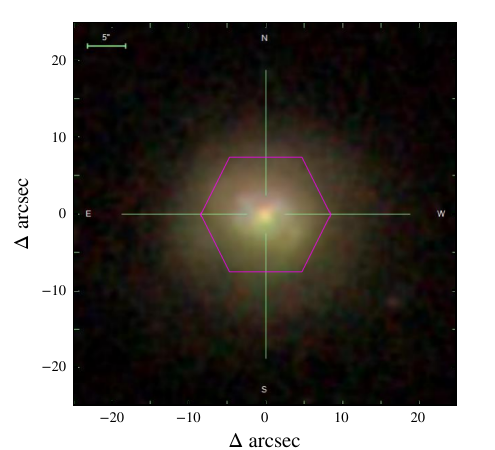}
\caption{The SDSS \citep{York_2000} false-color ($g$-, $r$-, and $i$-bands) image of PGC 37535. The magenta
hexagon shows the MaNGA bundle allocation.
\label{fig:optical_img}}
\end{figure}

PGC39535 is a massive star-forming S0 locating at the center of a massive group. It is a promising candidate of S0s which contains cool gas. Figure \ref{fig:optical_img} shows the Sloan Digital Sky Survey \citep[SDSS;][]{York_2000} image of PGC 39535. We carry out CO observation for PGC 39535 with Northern Extended Millimeter Array (NOEMA) to investigate the molecular gas properties. Table \ref{tab:prop} summarizes the global properties of this galaxy from previous studies and this work. 

The paper is organized as follows. In Section \ref{sec:manga_data} and \ref{sec:noema}, we analyze its stellar and gas components properties using Mapping Nearby Galaxies at APO (MaNGA) integral field spectroscopy and NOEMA observations, respectively. In Section \ref{sec:discussion}, we discuss the gas distribution, molecular gas main sequence (MGMS), Kennicutt-Schmidt (K-S) law and star formation history derived by the combination of optical and sub-millimeter data. In Section \ref{sec:summary}, we present the summary. Throughout the paper, we adopt a $\Lambda$CDM cosmology with $\Omega_{\mathrm{M}}=0.3, \Omega_{\Lambda}=0.7$ and $H_0=70 \mathrm{~km} \mathrm{~s}^{-1} \mathrm{Mpc}^{-1}$, and a Salpeter initial mass function \citep[IMF;][]{Salpeter_1955}.

\begin{table}[h!]
  \begin{center}
    \caption{Global properties of PGC 39535}
    \label{tab:prop}
    \begin{tabular}{lcc}
      \toprule 
      \textbf{Quantity} & \textbf{Value} & \textbf{Reference}\\
      \midrule 
      R.A.(J2000) & 184.62321 & SDSS\\
      Decl.(J2000) & 35.622257 & SDSS\\
      Redshift & 0.0348965 & SDSS\\
      log $M_*$(M$_\odot$) & 10.70 & MPA-JHU$^{a}$\\
      SFR(M$_\odot$yr$^{-1}$) & 1.57 & this work\\
      log $\Sigma_{\rm SFR}$(M$_\odot$yr$^{-1}$kpc$^{-2}$) & -1.54 & this work\\ 
      log $M_{\rm H_{2}}$(M$_\odot$) & 9.73 $\pm$ 0.12 & this work\\
      log $\Sigma_{\rm H_{2}}$(M$_\odot$pc$^{-2}$) & 2.13 $\pm$ 0.12 & this work\\
      log $\Sigma_{\rm gas}$(M$_\odot$pc$^{-2}$) & 2.16 $\pm$ 0.26 & this work\\  
      \bottomrule 
    \end{tabular}
  \end{center}
  \begin{tablenotes}
  \small
  \item The M$_{\rm H_2}$ and $\Sigma_{\rm gas}$ are derived within the CO region in Figure \ref{fig:overlap}. The superscript $^{a}$ indicates the value is from MPA-JHU catalog \citep{Kauffmann_2003b,Salim_2007}. The stellar mass and star formation rate are converted to a Salpeter IMF \citep{Salpeter_1955}.
  \end{tablenotes}
\end{table}

\section{MaNGA Optical Data} \label{sec:manga_data}

\subsection{Optical MaNGA Data}

Mapping Nearby Galaxies at APO (MaNGA) is one of the key projects of the SDSS-IV survey \citep{Blanton_2017}, with the objective of meticulously mapping the components and structures of approximately 10,000 nearby galaxies. These observations are conducted using the Sloan 2.5m telescope \citep{Gunn_2006} equipped with the two dual-channel Baryon Oscillation Spectroscopic
Survey (BOSS) spectrographs \citep{Smee_2013}. MaNGA utilizes individual optical fibers that are grouped into bundles of varying sizes, arranged in a closely packed, hexagonal formation known as integral field units (IFUs). The observed wavelength spans the range from 3600$\AA$ to 10400$\AA$, with a resolution of approximately 2000 \citep{Drory_2015}. The target galaxies in the primary samples were observed within 1.5 times their effective radius.
The data analysis pipeline \citep[DAP;][]{Belfiore_2019,Westfall_2019} produces derived data products, including spatially resolved maps of emission lines, such as H$\alpha$, H$\beta$, [O\textsc{iii}] and [N\textsc{ii}]. The spectral indices, such as Dn4000, Mgb, Fe5270 and Fe5335, and stellar and gas kinematics used in this work are also from the DAP file. Additionally, the Pipe3d pipeline generates data products containing information about the galaxy's stellar population, such as spatially resolved stellar masses, which are available in the Pipe3d value-added catalogs \citep{Sanchez_2016}.
Since PGC 39535 is one of the target galaxies in the primary samples, we utilize the DAP and Pipe3d value-added catalogs to investigate its stellar and ionized gas properties.

\subsection{MaNGA Data Analysis}
\begin{figure*}[!ht]
\includegraphics{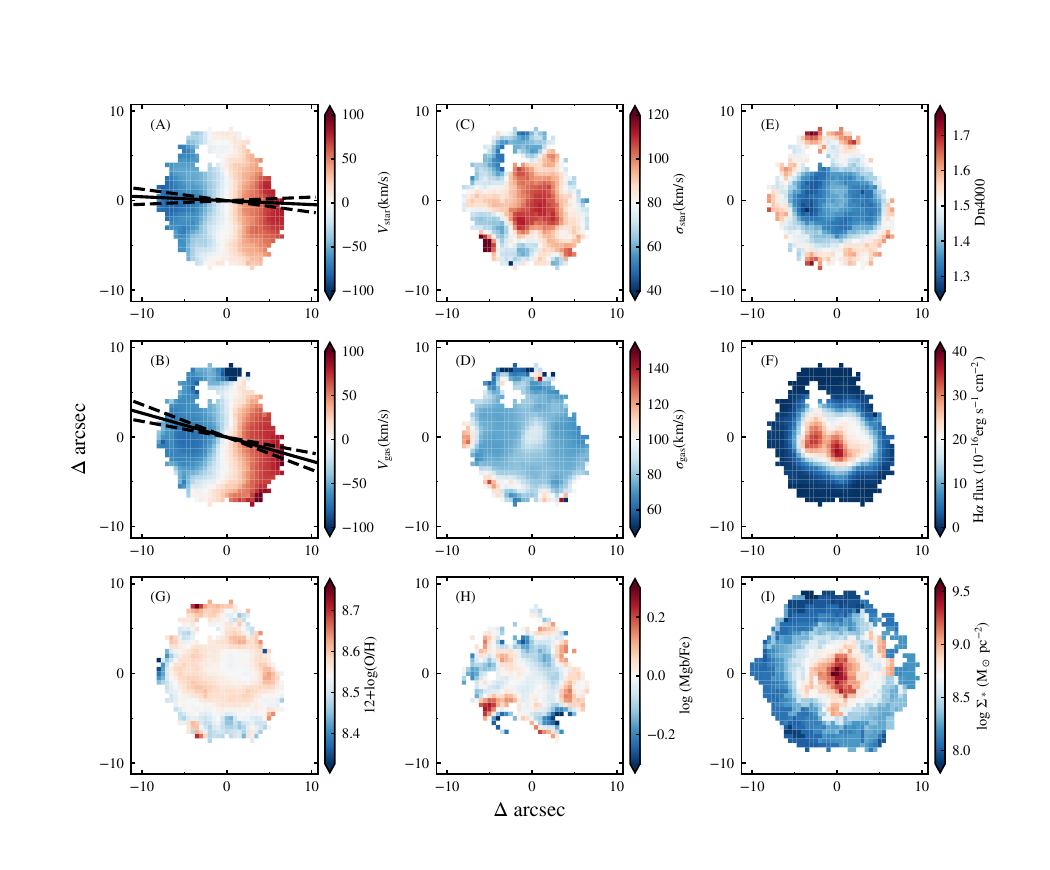}
\caption{Distributions of physical parameters color coded by the stellar velocity (Panel (A)), the gas velocity (Panel (B)), the stellar velocity dispersion (Panel (C)), the gas velocity dispersion (Panel (D)), Dn4000 (Panel (E)), the corrected H$\alpha$ flux density (Panel (F)), gas-phase metallicity (Panel (G)), the ratio of Mgb to Fe (Panel (H)), the the stellar mass surface density (Panel (I)), respectively. The solid black lines and dashed lines in panel (A) and (B) shows the major axes of the velocity field and the corresponding 1 $\sigma$ errors fitted by \textsc{python} module FIT\_KINEMATIC\_PA \citep{Kaviraj_2007}.
\label{fig:manga_map}}
\end{figure*}

The velocity and velocity dispersion of stellar and ionized gas components are depicted in Panels (A), (B), (C), and (D) of Figure \ref{fig:manga_map}, respectively.
Both the stellar and gaseous components exhibit regular rotation in their kinematics. We employ \textsc{python} module FIT\_KINEMATIC\_PA \citep{Kaviraj_2007} to fit the position angles (PA) of the stellar and gaseous components, and obtain PA$_{\rm star}=$267\degree.5$\pm$4\degree.8 and PA$_{\rm gas}=$255\degree$\pm$5\degree, respectively. The disks of stellar and ionized gas are kinematically aligned with each other, while a slight twist in the south of the ionized gas may be attributed to gas flows or stellar feedback as revealed by \cite{Roberts-Borsani_2020}. The velocity dispersion map of ionized gas also shows some enhancements at the outer part.
Panel (E) displays the distribution of Dn4000, indicating that the stellar populations in the outer regions of the galaxy are older compared to those in the inner parts. There are clusters of younger populations arranged in a ring-like structure surrounding the core region, implying recent active star formation.
The H$\alpha$ flux is corrected using the \citet{Calzetti_2000} extinction law, and the resulting corrected map is presented in Panel (F).

The gas-phase metallicity shown in Panel (G) is measured based on the O3N2 method according to \cite{Marino_2013},
\begin{equation}
12+\log (\mathrm{O} / \mathrm{H})=8.533-0.214 \times \mathrm{O} 3 \mathrm{N} 2
\end{equation}
where $\mathrm{O} 3\mathrm{N} 2=\log \left(\frac{[\mathrm{O}\textsc{iii}] \lambda 5007}{\mathrm{H} \beta} \times \frac{\mathrm{H} \alpha}{[\mathrm{N}\textsc{ii}] \lambda 6583}\right)$.
The oxygen abundance in galaxies is modified by multiple processes. Apart from the production and consumption processes through star formation and stellar remnants inside the galaxy \citep{Pagel_1975}, interactions with the environment can also arrange the metal contents, such as galaxy winds \citep{Heckman_1990}, accretion \citep{Keres_2005}, and mergers \citep{Rupke_2010}.
Most local galaxies show a negative metal gradient \citep[e.g.,][]{Searle_1971,Shaver_1983}, which is consistent with the inside-out formation scenario. Tidal-induced inflow through mergers can dilute the central metallicity and invert the gradient \citep[e.g.,][]{Kewley_2006,Ellison_2008}. On the other hand, merger-induced star formation can enrich the metallicity. \citet{Torrey_2012} found that the star formation effect overwhelms the inflowing gas effect and enhances the galaxy's central metallicity in gas-rich disk-disk mergers through numerical simulations.
The gas-phase metallicity in PGC 39535 exhibits higher values around the central region with strong star-forming activities, indicating the influence of star formation feedback.

The Mgb/$\rm \left \langle Fe \right \rangle$ ratio is defined as Mgb/$\rm \left \langle Fe \right \rangle$=\0Mgb/0.5(Fe5270+Fe5335)，which serves as a proxy for $\alpha$/Fe abundance. It traces the relative strength of intense starburst compared to long-term star formation \citep[e.g.,][]{Matteucci_1986,Thomas_2005}. The spatially irregular distribution of Mgb/$\rm \left \langle Fe \right \rangle$ in Panel (H) indicates the diversity of stellar population.
The stellar mass surface density of PGC 39535 in the central region displays a higher value as shown in Panel (I).

\begin{figure*}
\includegraphics{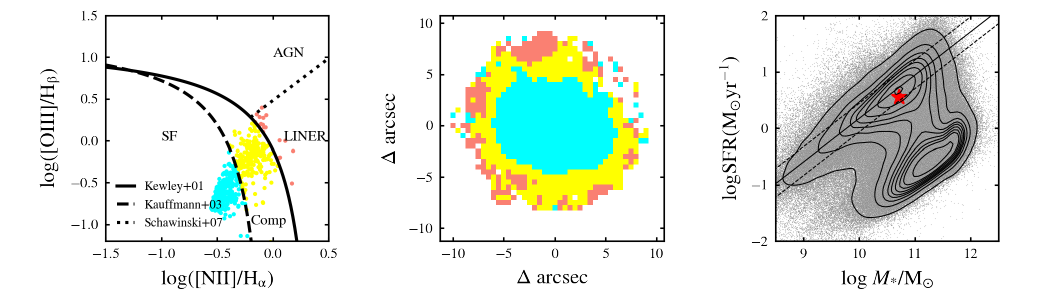}
\caption{The properties regarding star formation of PGC 39535. The left and middle panel shows the BPT diagram and BPT diagram map. The dashed, solid and dotted lines are the boundaries of star formation, composite, LINERs and AGN regions respectively \citep{Kewley_2001,Kauffmann_2003a,Schawinski_2007}. The cyan, yellow and salmon color represents SF, composite and LINERs regions respectively. The right panel shows the location of PGC 39535 on the SFMS, which is denoted by a red star. The SFMS and relative uncertainty are derived by \citet{Elbaz_2007}, shown as the solid and dashed lines. The background contour and scatters are from SDSS DR7 \citep{York_2000} galaxies.
\label{fig:sf}}
\end{figure*}

The spatially resolved and global information regarding star formation is depicted in Figure \ref{fig:sf}. We utilize the BPT diagram, [O\textsc{iii}]$\lambda$5007/H$\beta$ vs. [N\textsc{ii}]$\lambda$6583/H$\alpha$, to determine the excitation mechanism of each spaxel. We divide PGC 39535 into four regions: star-forming, composite, low-ionization
nuclear emission regions (LINERs), and active galactic nuclei (AGN) regions following \citet{Kewley_2001}, \citet{Kauffmann_2003a} and \citet{Schawinski_2007}.
The majority of spaxels in PGC 39535 belong to star-forming regions, and concentrate in the central region, which aligns with the stellar population properties depicted in Figure \ref{fig:manga_map}. The outskirts are predominantly composed of composite and LINER regions. The right panel of Figure \ref{fig:sf} shows the location of PGC 39535 relative to star-forming main sequence (SFMS). The SFMS is derived by \cite{Elbaz_2007}, shown as the black lines. By obtaining the stellar mass and total star formation rate (SFR) from the MPA-JHU catalog \citep{Kauffmann_2003b,Brinchmann_2004,Salim_2007}, we find that PGC 39535 follows the star-forming main sequence (SFMS) within the error.

\section{NOEMA Data} \label{sec:noema}
\subsection{NOEMA Millimeter Observation}
PGC 39535 was observed 3 times on 2023 March 15th, 19th and 23rd, respectively (Project W22CR; PI: Shiying Lu).
The observation was carried out with 12 antennas in the B configuration, and the total on-source exposure time is around 6.8 hours.
During the first 1.9h observation (on March 19th), 1156+295 was used as the receiver bandpass (RF), 1156+295 and 1144+402 as the phase/amplitude calibrators, and 0923+392 as the absolute flux calibrator.
On March 19th, 1156+295 was used as the RF, phase/amplitude and absolute flux calibrator simutaneously.
During the last 3.8h observation (on March 23rd), 3C84 was used as the RF calibrator, while 1156+295 and LKHA101 as the phase/amplitude and absolute flux calibrator, respectively.
The antennas are equipped with dual polarizations and each polarization covers a bandwidth of $\sim$4.1 GHz at a spectral resolution of 2MHz.

The calibration was performed using Continuum and Line Interferometer Calibration (CLIC), and the cleaning and imaging process were performed using MAPPING. Both CLIC and MAPPING are the modules in the software package of GILDAS\footnote{\url{https://www.iram.fr/IRAMFR/GILDAS}}.
The redshifted CO(1-0) frequency is 111.38 GHz at given $z$=0.0349 (the rest-frame frequency $\nu_{\rm rest}$=115.27 GHz).
We image a CO(1-0) datacube which consists of 640$\times$640  pixels for a map cell of 0$\arcsec$.2$\times$0$\arcsec$.2, covering a field of view (FoV) of 47$\arcsec$.3$\times$47$\arcsec$.3. at a velocity resolution of 10 km/s. The CO flux is in the unit of Jy/beam.
The synthesized beam size is 1.39$\arcsec$$\times$0.89$\arcsec$ with PA=157\degree.5.

\subsection{CO Distribution}
\label{sec:co_distribution}
\begin{figure*}[!ht]
\includegraphics{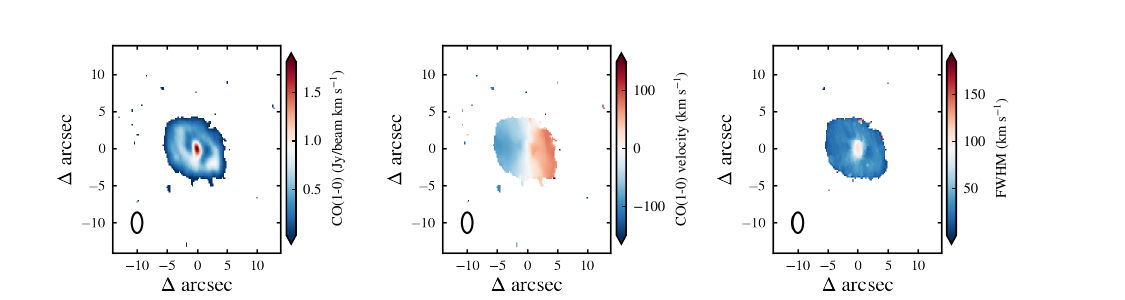}
\caption{CO(1-0) moment maps of PGC 39535. The moment maps are generated using a 2.65 mJy ($\sim$ 5 $\sigma_{rms}$) threshold. From left to the right, it shows the CO intensity, velocity and velocity dispersion, respectively. The beam size is shown at the bottom-left corner in each panel.
\label{fig:co_distribution}}
\end{figure*}

\begin{figure}
\includegraphics{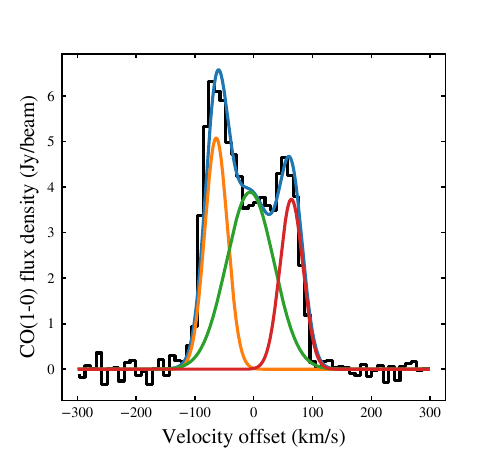}
\caption{The integral flux spectrum of CO(1-0). The observed integral flux distribution is represented by the black line with a velocity interval of 10 km/s. The flux profile is fitted using three Gaussian components. The blue line corresponds to the superposition of these three Gaussian components, representing the best-fit model.
\label{fig:integ_flux}}
\end{figure}

\begin{figure*}
\includegraphics{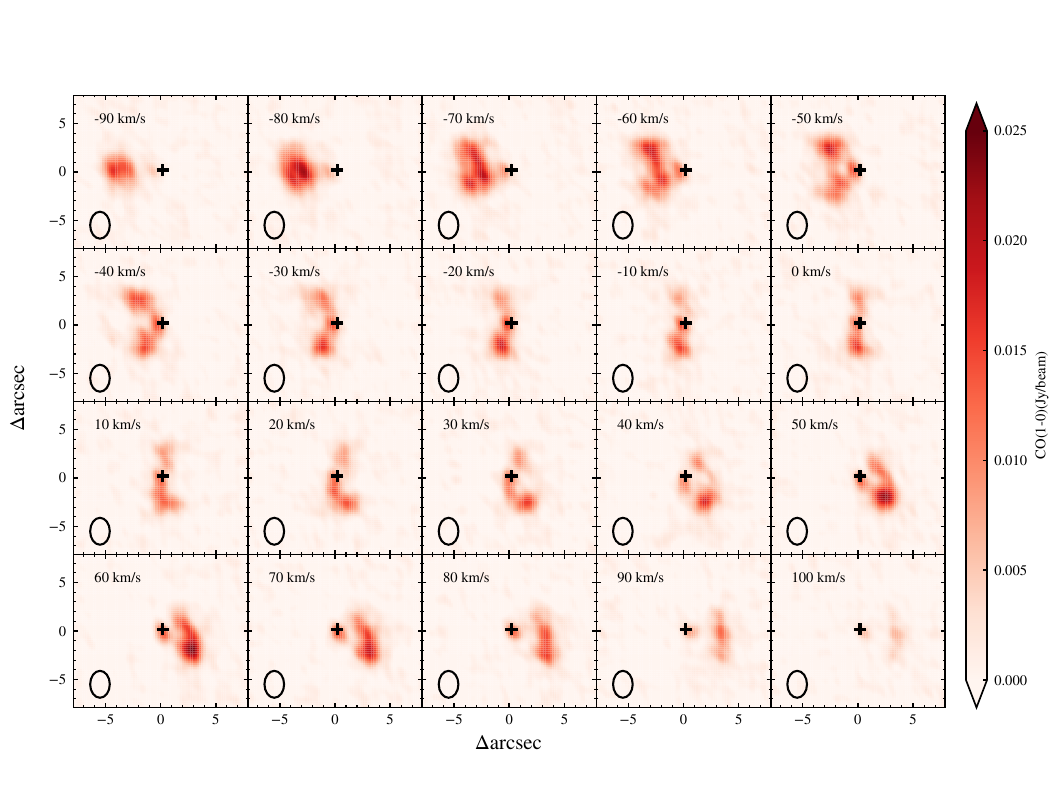}
\caption{The channel maps of CO(1-0) with a velocity range of -90 $\rm km\; s^{-1}$\textless v\textless 100 $\rm km\; s^{-1}$ and $\Delta$v=10 $\rm km s^{-1}$. The scale is consistent across each panel, and the color bar showing flux density level is displayed on the right side. In each panel, the beam size is shown in the bottom left corner, while the corresponding velocity is shown in the top right corner. The center of galaxy is denoted by a black square.
\label{fig:channel_map}}
\end{figure*}

We apply a threshold of 2.65 mJy ($\sim$ 5 $\sigma_{\rm rms}$) to generate the CO moment maps. Figure \ref{fig:co_distribution} illustrates the distribution of CO intensity, velocity, and velocity dispersion from left to right, respectively. The majority of CO is concentrated in the central ~10$\times$10 arcsec$^2$ region. The CO distribution exhibits a peak at the center and forms a ring-like structure around it. Notably, the central region and the ring-like structure are connected through a molecular bar located in the southern part of the galaxy, which may facilitate gas exchange, as will be discussed in Section \ref{sec:gas_inflow}. The velocity field of CO demonstrates a regular rotation disk. Using FIT\_KINEMATIC\_PA module of \textsc{python} \citep{Kaviraj_2007}, we measure the kinematic position angle of CO (PA$_{\rm CO}$=252\degree.5$\pm$0\degree.5), which aligns with the position angles of the stellar (PA$_{\rm star}$=267\degree.5$\pm$4\degree.8) and ionized gas components (PA$_{\rm gas}$=255\degree$\pm$5\degree). The CO velocity dispersion increases from the outer regions towards the center.
To obtain an integral spectrum of CO, we combine the spectra of each pixel with CO detection, as shown in Figure \ref{fig:co_distribution}. We employ double narrow Guassian profiles to reproduce the rotation disk component, and a broad Gaussian profile
to generate the central peak component.
In Figure \ref{fig:integ_flux}, the orange and red lines are representatives of the approaching and the receding side in the rotation disk, while the broad green line are for the central peak component.
The structure of CO(1-0) becomes more evident in the channel map as shown in Fig \ref{fig:channel_map}, which encompasses a velocity range from -100 km/s to 100 km/s with a velocity interval of 10 km/s. In most channels, three main components are identified: the northern part and southern part of the ring-like structure, along with the central peak region.

We measure the CO luminosity from the integrated spectrum following \citet{Solomon_2005},
\begin{equation}
L_{\mathrm{CO}}^{\prime}\left[\mathrm{K}\; \mathrm{km} \mathrm{s}^{-1}\right]=3.25 \times 10^7 S_{\mathrm{CO}} \Delta v \nu_{\mathrm{obs}}^{-2} D_L^2(1+z)^{-3},
\end{equation}
where $S_{\mathrm{CO}} \Delta v$ is the integrated flux density in unit of Jy km/s, $\nu_{\mathrm{obs}}$ is the observed frequency of CO in gigahertz, and $D_L$=153.46 Mpc is the luminosity distance of PGC 39535. The uncertainty is defined as 
\begin{equation}
\delta I_{\mathrm{CO}}=\sigma_{\mathrm{RMS}}\left(W_{\mathrm{CO}} \delta v_c\right)^{1 / 2},
\end{equation}
where $\sigma_{\mathrm{RMS}}$ is the root-mean-square (RMS) noise in Jy, $W_{\mathrm{CO}}$ is the CO line width in km s$^{-1}$, and $\delta v_c$ is the spectral resolution \citep{Salvestrini_2022}.
Following the definition above, the CO luminosity is measured to be 1.26$\times$10$^9$ K km s$^{-1}$ pc$^{2}$. 
The corresponding H$_2$ mass is estimated using a conversion factor between CO and H$_2$:
\begin{equation}
\mathrm{M}_{\mathrm{H}_2}=\alpha_{\mathrm{CO}} \times L_{\mathrm{CO}}^{\prime},
\end{equation}
where $\alpha_{\rm CO}$ is taken to be the same as the inner disk of Milky Way $\alpha_{\mathrm{CO}} = 4.3\left(\mathrm{M}_{\odot} \mathrm{km} \mathrm{s}^{-1} \mathrm{pc}^2\right)^{-1}$ \citep{Bolatto_2013}. The molecular gas mass is estimated to be (5.42$\pm$1.52)$\times$10$^{9}$ M$_{\odot}$ considering the 10\% flux uncertainty of NOEMA at 3mm. 
The inclination of the galaxy is estimated from its ellipticity,
\begin{equation}
\cos i=\sqrt{\frac{(1-\epsilon)^2-q_0^2}{1-q_0^2}}
\end{equation}
where $i$ is the inclination angle, $\epsilon$ is the ellipticity and $q_0$ is taken to be 0.2 \citep{Tully_2000}. 
The inclination-corrected mass surface density of CO is estimated to be (1.36$\pm$0.38)$\times$10$^{2}$ M$_{\odot}$ pc$^{-2}$.
We calculate the star formation efficiency (SFE) by the ratio of SFR and molecular gas mass within their common area shown as the outermost countour indicated in Figure \ref{fig:overlap}.
The SFE is then estimated to be 2.12$\times$10$^{-10}$ yr$^{-1}$. The estimated SFE is similar with that of early-type galaxies (ETGs) \citep[4$\times$10$^{-10}$ yr$^{-1}$,][]{Davis_2015} and less than that of spirals \citep[1.5$\times$10$^{-9}$ yr$^{-1}$,][]{Kennicutt_1998}. The depletion time is estimated to be 1.5 Gyr assuming a constant gas consumption.

\section{Discussions} \label{sec:discussion}

\subsection{MGMS and K-S Law}
\label{KS}
\begin{figure}
\includegraphics{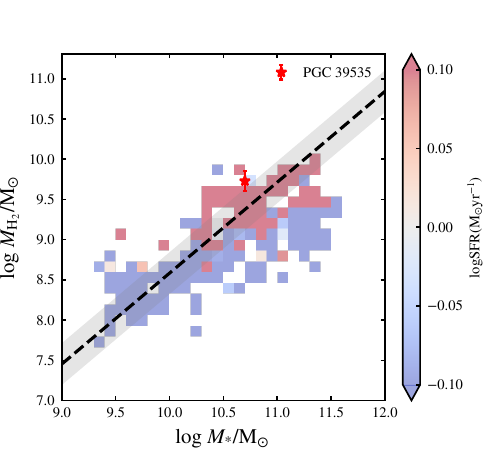}
\caption{The distribution of log $M_*$ vs. log $M_{\rm H_2}$, where PGC 39535 is shown as the red star. The mosaic distributions are galaxies in extended CO Legacy Database for GASS \citep[xCOLD-GASS;][]{Saintonge_2017} color-coded by their star formation rates. The molecular gas main sequence (MGMS) with 1$\sigma$ scatter shown in black solid line and gray shadow is derived in \citet{Baker_2023}.
\label{fig:MGMS}}
\end{figure}

Previous studies have demonstrated a positive correlation between $\Sigma_{\rm H_2}$ and $\Sigma_*$ \citep[e.g.][]{Wong_2013,Barrera-Ballesteros_2018}. It is believed that the stellar mass reflects the underlying gravitational potential, where the molecular gas is either retained or formed due to higher pressure \citep{Lin_2019a}. This so-called molecular gas main sequence (MGMS) is studied both globally and spatially resolved, and is considered more fundamental than the SFMS \citep[e.g.,][]{Lin_2019a,Ellison_2021,Baker_2023}.
The MGMS is depicted in Figure \ref{fig:MGMS}, where the red star represents PGC 39535. The background color-coded values are normal galaxies representative of local galaxy population, obtained from extended CO Legacy Database for GASS (xCOLD GASS) \citep{Saintonge_2017}. Galaxies with higher SFRs tend to be situated at the upper end of MGMS, indicating that star formation is intensified in galaxies with an increased gas fraction. PGC 39535 exhibits a slightly enhanced gas fraction compared to MGMS.

\begin{figure}
\includegraphics{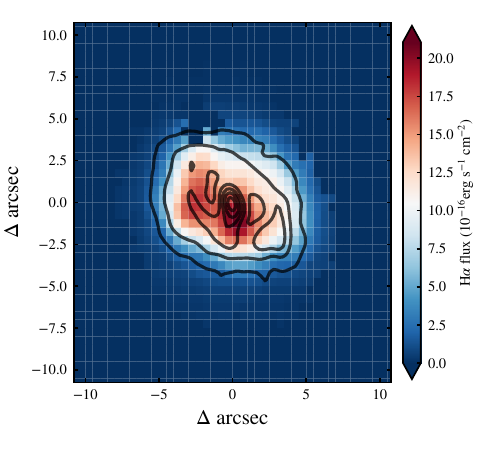}
\caption{The distribution of H$\alpha$ and CO(1-0) flux in PGC 39535. The contours represent the CO(1-0) intensity, and the outermost contour corresponds to 0.03 Jy/beam km s$^{-1}$ level with the increments of 0.3 Jy/beam km s$^{-1}$ from outside-in. The background mosaic is the corrected H$\alpha$ intensity, with the colorbar shown at the right side.
\label{fig:overlap}}
\end{figure}

In Figure \ref{fig:overlap}, it shows the corrected H$\alpha$ flux distribution, overlaid on the CO(1-0) contours with the increments of 0.3 Jy/beam km s$^{-1}$,  starting from the value of 0.03 Jy/beam km s$^{-1}$ at the outermost. The H$\alpha$ distribution shows two clumps: the first clump is situated near the center, slightly offset towards the junction between the central molecular bar and ring, while the second clump is found within the ring. The gas flowing along the molecular bar (as will be discussed in Section \ref{sec:gas_inflow}) could increase the gas surface density and enhance the star formation activity. Gas is the raw material of star formation. The molecular and ionized gas, traced by CO(1-0) and H$\alpha$ emission, exhibit similar distributions in PGC 39535. The region within the outermost contour of CO in Figure \ref{fig:overlap} is selected in this work to calculate both gas mass surface density ($\Sigma_{\rm gas}$) and star formation rate surface density ($\Sigma_{\rm SFR}$).
Following the method described in Sec \ref{sec:co_distribution}, the molecular gas mass surface density is $\Sigma_{\rm CO}$=(1.36$\pm$0.38)$\times$10$^{2}$ M$_{\odot}$ pc$^{-2}$.
Due to the lack of spatially-resolved HI data, we employ an empirical conversion between HI mass surface density and metallicity to estimate this value \citep{Schruba_2018}.
The surface density of HI can be estimated using
\begin{equation}
\log _{10} \Sigma_{\mathrm{HI}}=(-0.86 \pm 0.19) \log _{10} Z^{\prime}+(0.98 \pm 0.04),
\end{equation}
where $Z^{\prime}$ is the metallicity normalized to solar neighborhood metallicity of $12+\log (\mathrm{O} / \mathrm{H})=8.76$.
As a result, the total gas mass surface density ($\Sigma_{\rm gas}$) in logarithmic space is estimated to be 2.16$\pm$0.26 M$_{\odot}$pc$^{-2}$, which is listed in Table \ref{tab:prop}.
The star formation rate density ($\Sigma_{\rm SFR}$) is derived by converting the H$\alpha$ flux within the same region following \citet{Kennicutt_1998}:
\begin{equation}
\mathrm{SFR}\left(\mathrm{M}_{\odot} \mathrm{yr}^{-1}\right)=7.9 \times 10^{-42} L(\mathrm{H} \alpha)\left[\mathrm{erg} \cdot \mathrm{s}^{-1}\right]
\end{equation}

\begin{figure}
\includegraphics{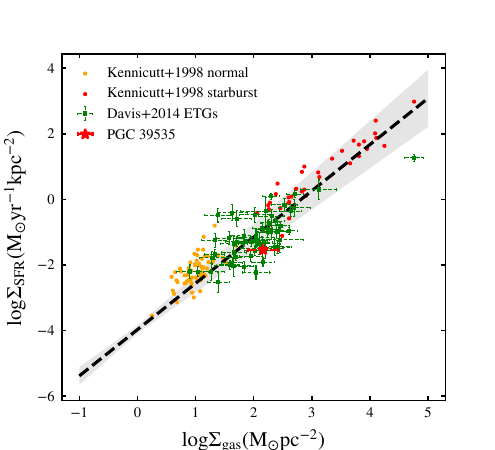}
\caption{The distribution of log $\Sigma_{\rm gas}$ vs. log $\Sigma_{\rm SFR}$ (i.e., K-S law). The red, orange and green points represent the distribution of starburst spirals, normal spirals from \citet{Kennicutt_1998} and ETGs from \cite{Davis_2014}, respectively. The dashed black line and shaded region represents the K-S law derived using normal spirals and starburst spirals from \citet{Kennicutt_1998}. PGC 39535 in this work is shown as the red star.}
\label{fig:ks_law}
\end{figure}

In Figure \ref{fig:ks_law}, we show the location of PGC 39535 (black star) relative to the Kennicutt-Schmidt law (i.e., K-S law), compared to starburst spirals \citep[red dots,][]{Kennicutt_1998}, normal spirals \citep[orange dots,][]{Kennicutt_1998}, and ETGs \citep[green squares,][]{Davis_2014}. We find that PGC 39535 falls slightly below the Kennicutt-Schmidt relation.
Our results are consistent with \cite{Davis_2014}, where they found a systematic offset of ETGs from the K-S relation. The lower star formation efficiency may be due to the shear motion caused by the deep gravity potential of the massive bulge.
The shear rate is directly related to $\beta$, which is defined as the logarithmic derivative of the rotation curve \citep{Davis_2014}. To obtain the rotation curve, we fit the 3D data cube with a tilted-ring model using $^{\rm 3D}$Barolo \citep{DiTeodoro_2015}, as will be described in Section \ref{sec:gas_inflow}. The average value of $\beta$ within the galaxy where the rotation cube reaches its peak, is $\sim$ 0.2. This value is between galaxies with the strongest suppresion of star formation ($\sim$ 0.1) and normal galaxies ($\sim$ 0.35) in \citet{Davis_2014}. Therefore, shear is one possible mechanism to suppress star formation in PGC 39535.

\subsection{Star Formation History}

Galaxies in massive halos follow an inside-out quenching scenario, which can be attributed to AGN feedback and/or morphology quenching \citep{Lin_2019b}. According to the halo mass catalog of \cite{Yang_2007}, we find that PGC 39535 is located at the center of a massive group with halo mass of 10$^{12.3}$ M$_{\odot}$ but it shows central star formation, which indicates that PGC 39535 may have undergone merger events, resulting in the gas inflow towards the center and subsequent enhancement of central star formation \citep[e.g.,][]{Barnes_1996,Bekki_2011,Ellison_2013,Moreno_2015}. Observations and simulations have shown that mergers can increase star formation in the center and suppress it at larger radii in a short period, followed by a rapid truncation once the gas is fully consumed. 
The gas fraction, stellar mass ratio and orbital parameters can affect the remnant of galaxy mergers \citep[e.g.,][]{Lagos_2018,Hopkins_2009,Robertson_2006,Springel_2005}. Gas-rich mergers may even enhance the angular momentum of galaxies \citep{Moreno_2015}, which may be the possible reason why PGC 39535 still retains a significant fraction of rotation and high star formation rate.

\begin{figure*}
\includegraphics{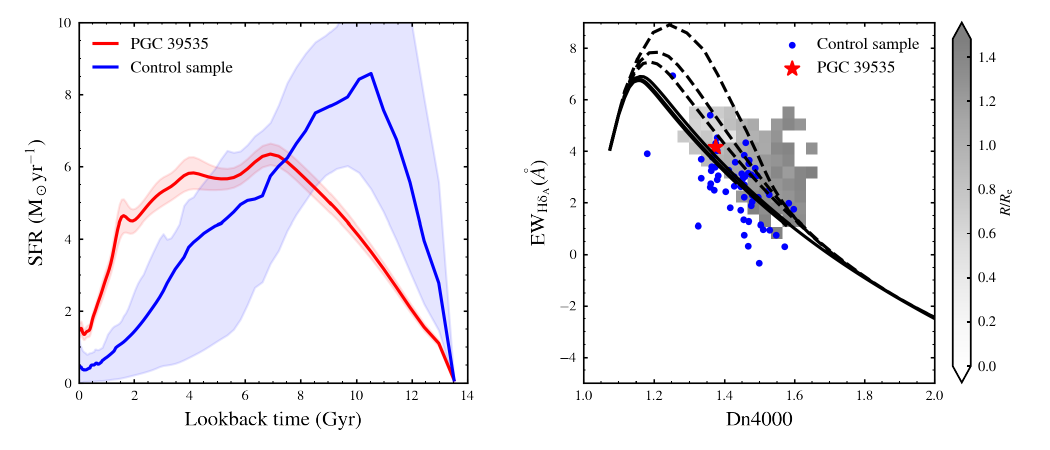}
\caption{The comparison of star formation history (left panel) and EW$_{\rm H\delta_A}$-Dn4000 distribution (right panel) between target PGC 39535 and the control sample. In the left panel, star formation histories of PGC 39535 and control sample are derived using the methodology from \cite{Gao_2024}. The red line and shaded region show the SFH and its error of PGC 39535 while the blue line and shaded region show the median SFH and its 16th to 84th percentile distribution of the control sample (57 galaxies) selected from SDSS. In the right panel, it shows the distribution of PGC 39535 (red star) and the control sample (blue dots) on the Dn4000 vs. EW$_{\rm H\delta_A}$ space. The gray-scaled mosaic indicates the radial location of the spatially resolved information of PGC 39535. The solid and dashed lines from bottom to the top are models generated under the assumption of exponentially declining star formation histories (i.e., $\tau$-model SFH$\sim$$e^{t/\tau}$) with characteristic declining timescales $\tau$ = 6, 4, 2, 0.6, 0.4, 0.2 Gyr.
\label{fig:sfh}}
\end{figure*}

Aiming to better reveal the evolution of PGC 39535, we try to derive the star formation history (SFH) by utilizing the methodology from \cite{Gao_2024}. The short description of this methodology is  as follows.
We employ the TNG100 hydrodynamical simulation \citep{Marinacci_2018,Naiman_2018,Nelson_2018,Nelson_2019,Pillepich_2018a,Pillepich_2018b, Springel_2018} as a prior and conduct spectral energy fittings (SEDs) on observed galaxies using four broad-band optical colors ($u$-$g$, $g$-$r$, $r$-$i$, $i$-$z$) from SDSS DR7 \citep{York_2000, Abazajian_2009}. The SFHs are then derived by building the mapping between color space and physical properties within simulations, and then applying it to real galaxies. The uncertainty in the SFH is quantified as the standard deviation of 50 bootstrap iterations. We smooth the SFHs and their error to avoid shot noise. The smoothing is applied with a Gaussian kernal with standard deviation of 2 Gyr. 
We select 57 galaxies as the control sample which satisfy the criteria of $\Delta$ log$M_*$/M$_{\odot}$\textless 0.1,  $\Delta$ log SFR\textless 0.2, $\Delta$ $z$\textless 0.015 from SDSS DR7, and assure their morpholigies of S0s using Galaxy Zoo project \citep{Willett_2013}. Only central galaxies are selected according to the group information in \cite{Yang_2007}. The medium SFH of control galaxies is also shown in Figure \ref{fig:sfh} by the blue line.
The SFR of the control sample decreases at a nearly constant pace after the enhancement phase (i.e., lookback time t\textless10 Gyr), while the declination phase of PGC 39535 breaks into two stages. The star formation declines slowly at the first 5 Gyr (i.e., lookback time 2Gyr\textless t\textless 7Gyr) then encountered a sharp decrease during the last 2 Gyr (i.e., lookback time t\textless 2 Gyr).

Compared to the control sample, the SFR in PGC 39535 exhibits a slower decline during the lookback time of 2-7 Gyr, which may be attributed to the enhancement of star formation caused by galaxy mergers, because mergers are more frequently found in dense (group) environments \citep[e.g.][]{Jian_2012}. The more rapid decrease in star formation over the last 2 Gyr may be a result of gas consumption following the mergers. However, the rate of change of the SFR slows down at $z=0$, consistent with the analysis in Section \ref{KS}, which may be caused by the low star formation efficiency.

Except for the SFH reconstruction, both Dn4000 and the equivalent width of Balmer absorption H$\delta_A$ line (i.e., EW(H$\delta_A$)) are used to constrain the ages of the stellar populations and the strength of bursty star formation over continuous star formation over the last few Gyr \citep[e.g.,][]{Kauffmann_2003b}. 
In Figure \ref{fig:sfh}, the global Dn4000 and EW(H$\delta_{\rm A}$) values for PGC 39535 and the control sample are obtained from the MPA-JHU catalog \citep{Kauffmann_2003b,Brinchmann_2004,Salim_2007}, while the spatially resolved values are from MaNGA DAP file \citep{Belfiore_2019,Westfall_2019}.
The spaxel distribution of PGC 39535 on the Dn4000 vs. EW(H$\delta_A$) diagram is shown by grey mosaics, scaled by the radial distance of spaxels (see the right panel of Figure \ref{fig:sfh}). Compared to the control sample, PGC 39535 has a higher EW(H$\beta_A$) and lower Dn4000, suggesting that it has a violent star formation over a short timescale, which is consistent with a merger scenario. The short timescale is more obvious on the spatially resolved scatters, where the closer to the center, the shorter characteristic declining timescale (i.e., $\tau$).

\begin{figure*}
\includegraphics{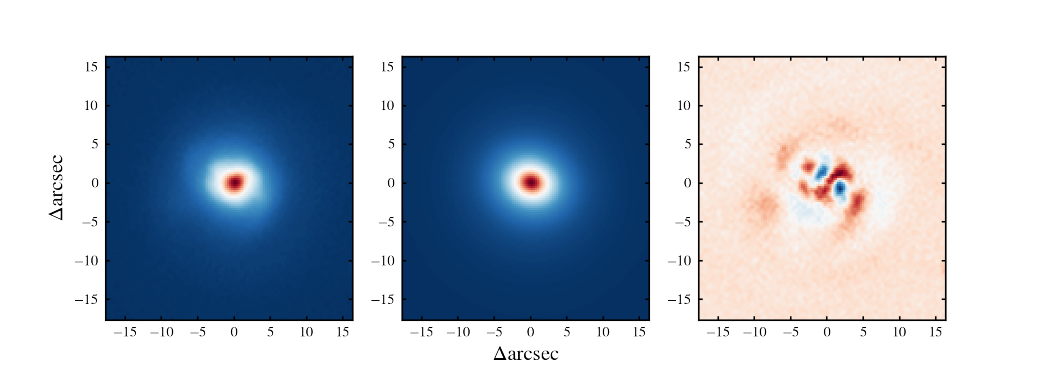}
\caption{The best-fit GALFIT \citep{Peng_2002,Peng_2010} results. The left, middle and right panels show the original r-band image, model generated by GALFIT and the residual image, respectively. The best-fit model is generated from GALFIT by adopting a S{\'e}rsic model plus an exponential profile. The residual is obtained by subtracting the best-fit model from the original image. 
\label{fig:galfit}}
\end{figure*}

However, the galaxy may have undergone rapid decline in asymmetry \citep[e.g.,][]{Lotz_2008} after the merger events, as there is no obvious asymmetry in the optical image.
We utilize GALFIT \citep[version 3.0.5;][]{Peng_2002,Peng_2010} to decompose surface brightness distribution of PGC 39535 by adopting a S{\'e}rsic and an exponential component. 
The S{\'e}rsic function has the following form:
\begin{equation}
\Sigma(r)=\Sigma_e \exp \left[-\kappa\left(\left(\frac{r}{r_e}\right)^{1 / n}-1\right)\right]
\end{equation}
where $\Sigma_e$ is the surface brightness at effective radius $r_e$, and $n$ is namely the S{\'e}rsic index. It indicates a flat core profile when the value is small. The exponential function is the special case of S{\'e}rsic profile when $n=1$.
The galaxy's $g$-band, $r$-band images and point spread function (PSF) are taken from DESI \citep{Dey_2019}. The parameters of the morphological components are presented in Table \ref{tab:galfit}. Figure \ref{fig:galfit} displays the original $r$-band image, model image, and residual image from left to right. The S{\'e}rsic index of bulge (i.e., $n_b$) suggests that PGC 39535 has a pseudo bulge and an exponential disk. The residual image reveals faint spiral-like structures, which may hint a faded spiral.
However, these structures are very faint which can not be found in the original optical image.

\begin{table}[h!]
  \begin{center}
    \caption{GALFIT fitting parameters}
    \label{tab:galfit}
    \begin{tabular}{lcccc}
      \toprule 
      \textbf{band} &\textbf{$n_b$}&\textbf{$r_b$}&\textbf{$r_d$}&\textbf{$B/D$}\\
      \textbf{(1)} &\textbf{(2)}&\textbf{(3)}&\textbf{(4)}&\textbf{(5)}\\
      \midrule 
      g & 0.69$\pm$0.01 & 2.97$\pm$0.01 & 4.44$\pm$0.01 & 0.58$\pm$ 0.02\\
      r & 0.98$\pm$0.01 & 3.29$\pm$0.01 & 4.86$\pm$0.02 & 0.82$\pm$ 0.03\\
      \bottomrule 
    \end{tabular}
  \end{center}
  \begin{tablenotes}
  \small
  \item (1)band, (2)S{\'e}rsic index of bulge, (3) bulge effective radius in arcsec, (4) disk scale length in arcsec, (5) bulge-to-disk light ratio.
  \end{tablenotes}
\end{table}

\subsection{Gas Inflow or Outflow}
\label{sec:gas_inflow}

\begin{figure}
\plotone{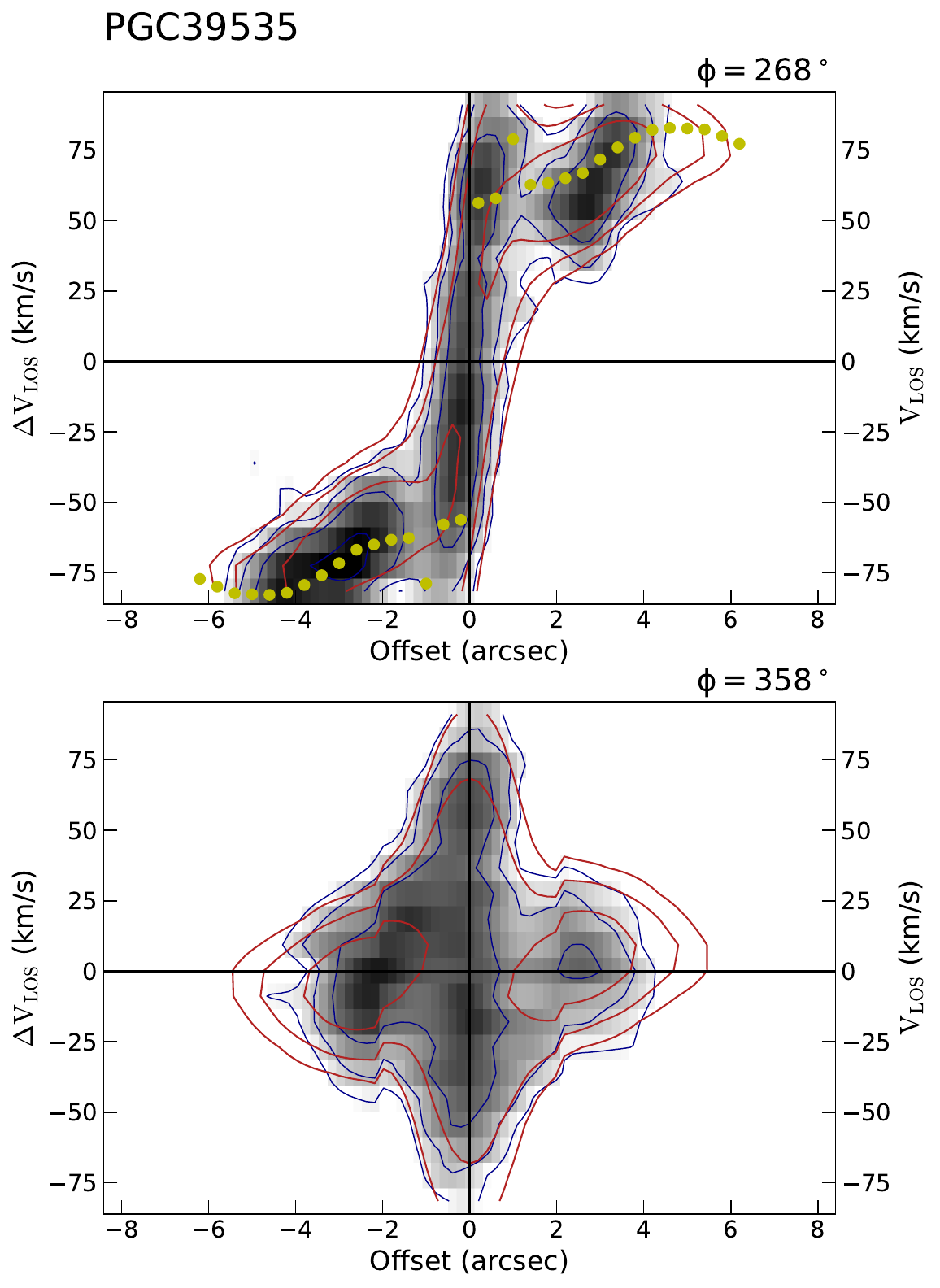}
\caption{The position-velocity diagram of PGC 39535, along the major (top panel) and minor (bottom panel) axis. The red contours are the best-fit model from $^{\rm 3D}$BAROLO \citep{DiTeodoro_2015}, while the blue contours and black mosaics represent the observed data. The dark yellow dots on the top panel present the projected rotation curve of the best-fit model.
\label{fig:pv_azim}}
\end{figure}

\begin{figure}
\plotone{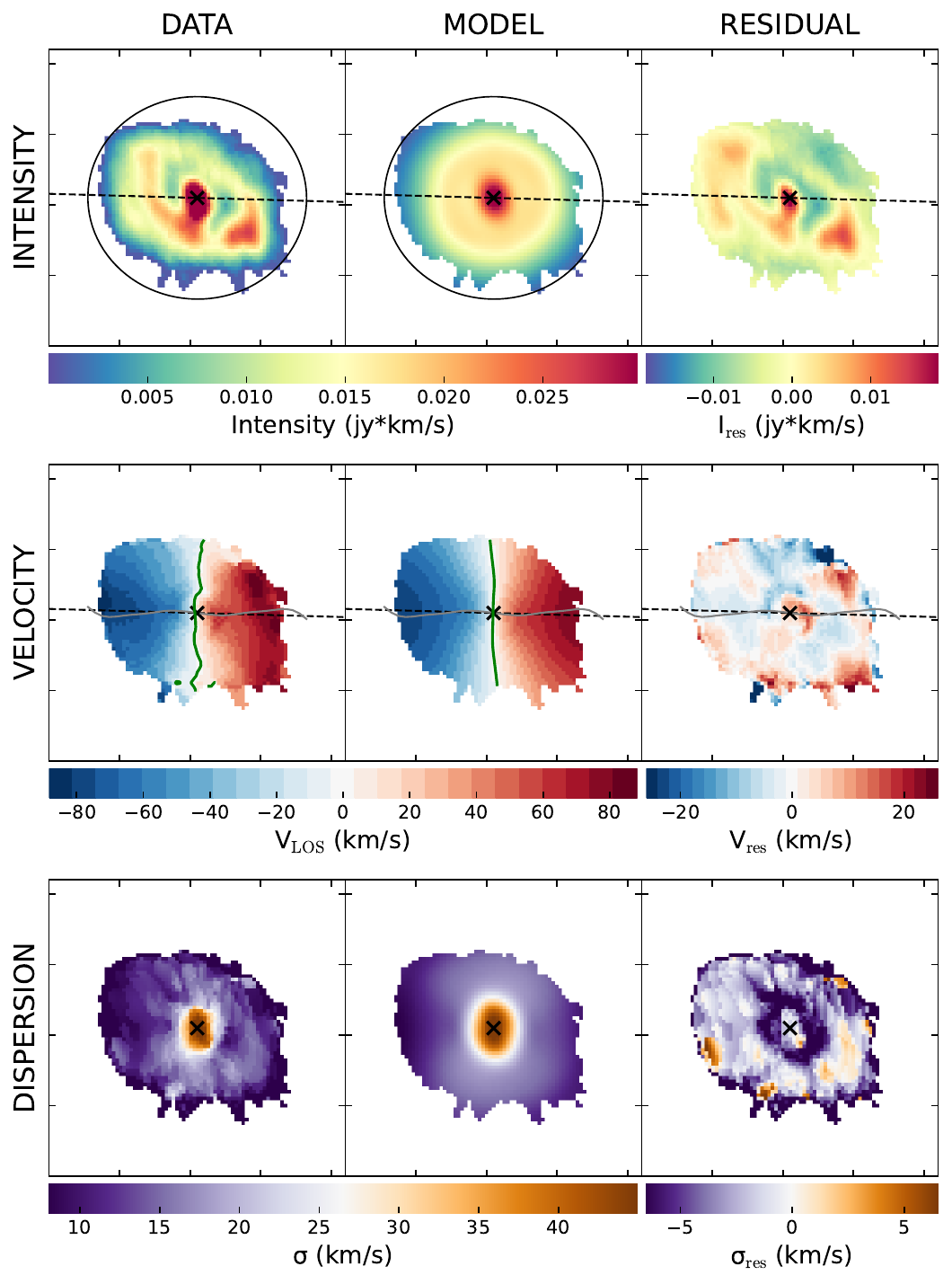}
\caption{The best-fit results of $^{\rm 3D}$BAROLO \citep{DiTeodoro_2015}. From top to bottom rows, it shows the intensity, velocity and velocity dispersion of moleclar gas CO(1-0) observation. From left to right columns, the original data, best-fit model, and residual are present, respectively. The model assumes titled-rings with circular rotation. The black circle on the top panels is the region used to fit. The major and minor axis of disk in the velocity map are shown in gray and green lines, separately. The galaxy center is marked by the black cross. 
\label{fig:maps_azim}}
\end{figure}

\begin{figure*}
\includegraphics{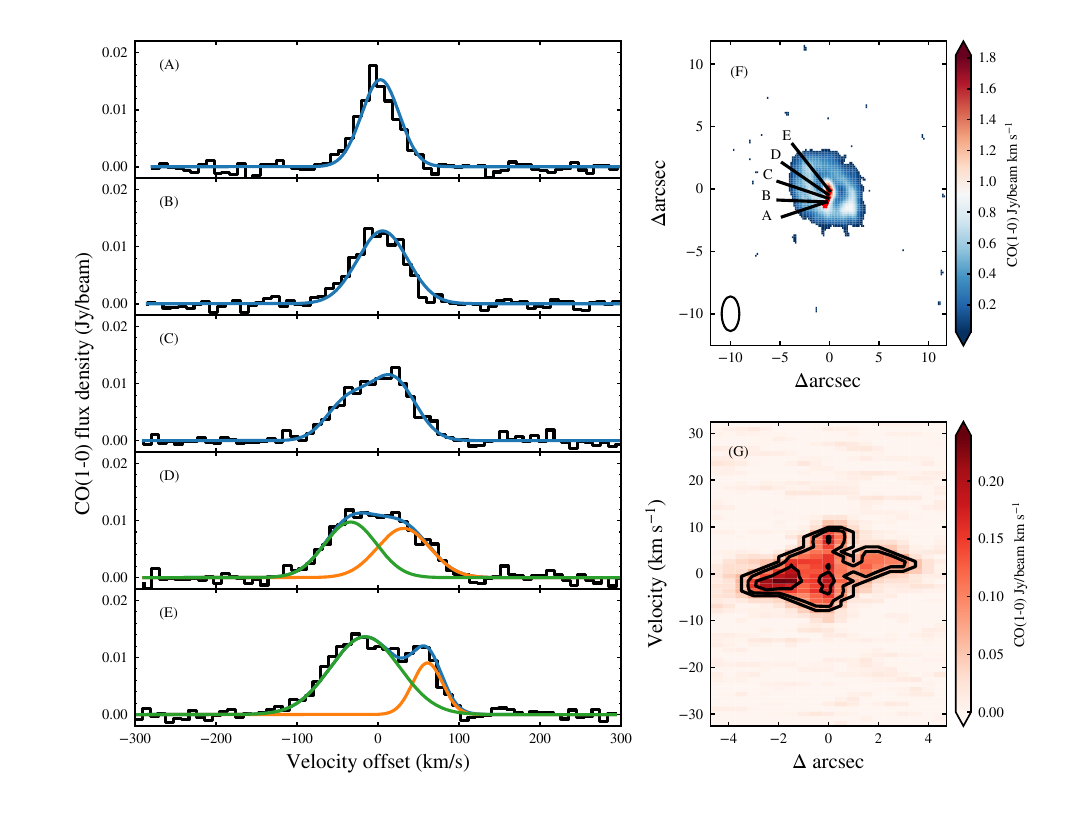}
\caption{The CO(1-0) flux distribution indicates gas flows along the molecuar gaseous bar. The left panel shows the spectrum extracted from the molecular band from south to the north. Each spectrum is fitted by a single Gaussian or a double Gaussian function. The corresponding pixel are marked by red stars in the upper right panel. The lower right panel shows the position-velocity diagram extracted from a pseudo slit along the minor axis with one arcsec width.
\label{fig:gas_exchange}}
\end{figure*}

In order to further study the kinematic and structures of PGC 39535, we employ the software $^{\rm 3D}$BAROLO provided by \cite{DiTeodoro_2015} to fit 3D tilted-ring models to CO emission-line datacubes, aiming to identify any gas motions deviating from regular rotation. The fitting process allows for the free adjustment of the initial parameters. We set the initial values of PA, inclination, rotation velocity, velocity dispersion as PA=261.47\degree, INC=17.74\degree, VROT=150 km/s, and VDISP=20 km/s.
Throughout the fitting process, the code iteratively fits a pure circular rotation model to the observed data. The resulting position-velocity diagram (PVD) is presented in Figure \ref{fig:pv_azim}. The upper panel displays the PVD along the major axis, while the lower panel shows the PVD along the minor axis. The red contours represent the best-fit model, while the blue contours and gray mosaics depict the observed data. Figure \ref{fig:maps_azim} showcases the original data, model, and residual images.
In general, the entire galaxy exhibits rotational motions, with the central molecular peak region and the ring-like structure standing out prominently. In the northern part of the galaxy, the disk component and central region are clearly distinct, whereas in the south, the velocity distribution spreads over a wider range, leading to a blending of the two parts as shown in the lower panel of \ref{fig:pv_azim}. The slight asymmetry observed in the position-velocity diagram along the minor axis and the residual velocity image in the lower panel of Figure \ref{fig:maps_azim} may indicate a gas exchange between these two regions. We generate the position-velocity diagram along a one-arcsec-width pseudo-slit along the minor axis in the Panel (G) of Figure \ref{fig:gas_exchange}, which indicates the asymmetry feature more clearly.

To make sure these patterns are real signal rather than noise caused by the imperfections of the model, we directly select five pixels along the molecular gaseous bar on the minor axis which connects the central peak region with the ring-like structure, and extract their corresponding spectra. As shown in Figure \ref{fig:gas_exchange}, we fit each spectrum with either a single Gaussian or two Gaussian components and observe a second component emerging from the outside-in.
The emergency of the second velocity component (shown in orange lines) indicates gas motion aside from regular rotation along the molecular bar. Gas exchange may potentially occur between the central region and the ring-like structure under the impact of gravitational potential and feedback.

Using stacking techniques of the Na D neutral gas tracer, \cite{Roberts-Borsani_2020} detected outflows in the centers of 78 galaxies from the MaNGA DR15 survey. The high $\Sigma_{\rm SFR}$ value and its correlation with mass outflow rates and loading factors indicate that star formation is the primary trigger for these outflows. In Figure \ref{fig:overlap}, the region within the outermost contour displays a log$\Sigma_{\rm SFR}$ of -1.54 M$_\odot$yr$^{-1}$kpc$^{-2}$ and log$\Sigma_*$ of 8.83$\pm$0.03 M$_\odot$kpc$^{-2}$, which corresponds to a neutral gas mass outflow rate log $\dot{M}_{out}$/M$_{\odot}$yr$^{-1}$ $\sim$ -0.6, as suggested by their statistical analysis.

Gas cycles in the center of galaxies have been identified in many studies. \cite{Armillotta_2019} find self-consistent gas cycle in the Central Molecular Zone (CMZ) through hydrodynamical simulations, where the gas flows towards the center along the galaxy bar, while gas outflow is triggered by stellar feedback. The star formation exhibit an oscillatory pattern due to this gas cycle. 
Our work presents a similar scenario, wherein inflowing gas has triggered star formation activity in the central region of PGC 39535, with subsequent feedback influencing the gas in return.

\begin{figure}
\includegraphics{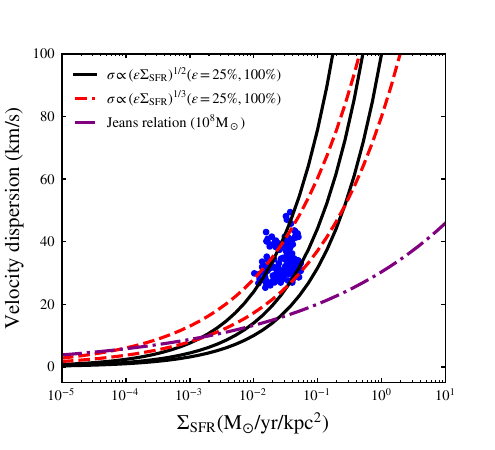}
\caption{Spatially resolved dependence of ionized gas velocity dispersion on star formation surface density of PGC 39535. The curves are models generated accroding to \cite{Lehnert_2009}. The solid lines represents models in the form of $\sigma \propto(\epsilon \dot{E})^{1 / 2}$, where $\dot{E}$ is the energy injection rate due to star formation and the $\epsilon$ is the coupling efficiency between injected energy and ISM. The dashed curves show models in the form of $\sigma_{\text {gas }} \propto(\epsilon \dot{E})^{1 / 3}$, where the energy is dissipated through turbulence. The purple line represents the gravity-powered turbulence model. The blue scatters are pixels unaffected by beam-smearing effects in PGC 39535 with SNR\textgreater 60.
\label{fig:turbulence}}
\end{figure}

The effect of star formation feedback is also reflected through its excitation of turbulent motion, which in turn plays an important role in regulating star formation \citep{Green_2010,Federrath_2012}.
In order to check the excitation mechanisms of  turbulent motions in the center galaxy of PGC 39535, we follow the methods of \cite{Zhou_2017} and select all spaxels in PGC 39535 with the observational signal-to-noise ratio (S/N) greater than 60. This enables us to resolve $\sigma_{\rm gas}$ down to 20 km/s, which is most cases in PGC 39535.
We correct all spaxels for the instrument resolutions, i.e., $\sigma_{\mathrm{gas}}=\left(\sigma_{\mathrm{obs}}^2-\sigma_{\mathrm{instr}}^2\right)^{1 / 2}$, where $\sigma_{\mathrm{instr}}$ is roughly 67 km/s. The velocity gradient for each spaxel is calculated using the upper, lower, left and right neighbors following equation (1) in \cite{Zhou_2017}. We then remove spaxels with v$_{\rm grad}$\textgreater 20 km/s and v$_{\rm grad}$\textgreater$\sigma_{\rm gas}$ to avoid the beam-smearing effects.
The relation between $\Sigma_{\rm SFR}$ and $\Sigma_{\rm gas}$ is depicted in Figure \ref{fig:turbulence}. After applying the S/N and beam-smearing cuts, the majority of the spaxels fall within a velocity dispersion range of 20-50 km/s and $\Sigma_{\rm SFR}$ range of $10^{-2}$-$10^{-1}$ M$_{\odot}$yr$^{-1}$kpc$^{-2}$.
The lines in Figure \ref{fig:turbulence} are different models proposed by \cite{Lehnert_2009} to explain the observed relation. We adopt the same parameter values as \cite{Zhou_2017} and summarize briefly as follows.
If star formation triggers the dissipation of energy, the relation would follow a simple form of $\sigma \propto(\epsilon \dot{E})^{1 / 2}$, where $\dot{E}$ is the energy injection rate due to star formation and the coupling efficiency between injected energy and ISM is represented by $\epsilon$. 
The bottom two solid black lines assume a conservative coupling efficiency of 25\%, while the top solid black line uses an extreme and unrealistic value of 100\%.
The two dashed curves show models in the form of $\sigma_{\text {gas }} \propto(\epsilon \dot{E})^{1 / 3}$, where the energy is dissipated through turbulence. The coupling efficiencies are taken to be 25\% and 100\% respectively. The purple line represents the gravity-powered turbulence model, which is the upper limit for velocity dispersion obtained using the Jeans relation. 
The majority of pixels in PGC 39535 are consistent with the $\sigma \propto(\epsilon \dot{E})^{1 / 2}$ and $\sigma \propto(\epsilon \dot{E})^{1 / 3}$ model, indicating the crucial role of star formation feedback in driving the turbulent motions in PGC 39535. However, most of the points are above the $\epsilon$=25\% model, which correspond to a more realistic case. 
Since PGC 39535 may suffer from shear as analyzed in Section \ref{KS}, it could be another source of turbulence driver \citep[e.g.][]{Krumholz_2017,Federrath_2016}. 
The gas accretion and outflow process may also influence the turbulent gas motions \citep{Glazebrook_2013}.

\section{Summary} \label{sec:summary}
Based on the 2D optical spectra from MaNGA and the millimeter observations from NOEMA, we conduct a study on a star-forming massive ($M_*$$\sim$5$\times10^{10}$ M$_\odot$) S0 galaxy PGC 39535. The availability of spatially resolved optical and millimeter data enables us to perform a comprehensive analysis of the stellar and gas components. In PGC 39535, star formation and gas primarily concentrate in the central region. Despite the abundance of molecular gas, the star formation efficiency is lower compared to star-forming spirals, possibly due to the central gravitational potential. Gas exchange occurs between different regions of PGC 39535, and the concentration of gas in the central region may be a result of a gas-rich merger event. The results of our investigation are presented below.
\begin{enumerate}
\item Analysis of the MaNGA 2D spectrum reveals that PGC 39535 exhibits a positive age-gradient. Both the stellar and gas components display regular rotation disks. The whole central region of the galaxy shows active star-forming activities, placing it on the star-forming main sequence (SFMS).
\item The NOEMA observations demonstrate that the molecular gas distribution overlaps with the star-forming region. The molecular gas mass is estimated to be (5.42$\pm$1.52)$\times$10$^{9}$ M$_{\odot}$. And the inclination-corrected mass surface density of CO(1-0) is estimated to be (1.36$\pm$0.38)$\times$10$^{2}$ M$_{\odot}$ pc$^{-2}$.
The distribution of molecular gas includes a rotational disk, a ring-like structure, and a central molecular peak region connected by a molecular bar. There is a possibility of gas exchange between the central molecular region and the ring-like structure according to the multi-Gaussian components fitting. The kinematics of CO(1-0) indicate a pattern of regular rotation.
\item PGC 39535 is located at the upper end of the molecular gas main sequence (MGMS), which contributes to its active star formation. Nevertheless, it exhibits a slight deviation from the Kennicutt-Schmidt (K-S) law, potentially due to the shear motion in the central region within the presence of a deep gravity potential. 
\item PGC 39535 is located in the center of a massive group with a mass of $10^{12.3}$ $M_{\odot}$. Based on the star formation history, we infer that the gas and star formation activity in the central region may have resulted from a gas-rich merger. Morphological decomposition reveals residual faint spiral structures, indicating its possible transition from a spiral galaxy.
\end{enumerate}

The authors are very grateful to the anonymous referee for
critical comments and instructive suggestions, which improved the content and analyses significantly in this work.
This work is supported by the National Natural Science Foundation of China (No.12192222, 12192220 and 12121003). This work is based on observations carried out with the IRAM Northern Extended Millimeter Array. IRAM is supported by INSU/CNRS (France), MPG (Germany), and IGN (Spain). In addition, we acknowledge the supports of the staff of NOEMA, especially Laure Bouscasse.

Funding for the Sloan Digital Sky 
Survey IV has been provided by the 
Alfred P. Sloan Foundation, the U.S. 
Department of Energy Office of 
Science, and the Participating 
Institutions. 

SDSS-IV acknowledges support and 
resources from the Center for High 
Performance Computing  at the 
University of Utah. The SDSS 
website is www.sdss4.org.

SDSS-IV is managed by the 
Astrophysical Research Consortium 
for the Participating Institutions 
of the SDSS Collaboration including 
the Brazilian Participation Group, 
the Carnegie Institution for Science, 
Carnegie Mellon University, Center for 
Astrophysics | Harvard \& 
Smithsonian, the Chilean Participation 
Group, the French Participation Group, 
Instituto de Astrof\'isica de 
Canarias, The Johns Hopkins 
University, Kavli Institute for the 
Physics and Mathematics of the 
Universe (IPMU) / University of 
Tokyo, the Korean Participation Group, 
Lawrence Berkeley National Laboratory, 
Leibniz Institut f\"ur Astrophysik 
Potsdam (AIP),  Max-Planck-Institut 
f\"ur Astronomie (MPIA Heidelberg), 
Max-Planck-Institut f\"ur 
Astrophysik (MPA Garching), 
Max-Planck-Institut f\"ur 
Extraterrestrische Physik (MPE), 
National Astronomical Observatories of 
China, New Mexico State University, 
New York University, University of 
Notre Dame, Observat\'ario 
Nacional / MCTI, The Ohio State 
University, Pennsylvania State 
University, Shanghai 
Astronomical Observatory, United 
Kingdom Participation Group, 
Universidad Nacional Aut\'onoma 
de M\'exico, University of Arizona, 
University of Colorado Boulder, 
University of Oxford, University of 
Portsmouth, University of Utah, 
University of Virginia, University 
of Washington, University of 
Wisconsin, Vanderbilt University, 
and Yale University.

This project makes use of the MaNGA-Pipe3D dataproducts. We thank the IA-UNAM MaNGA team for creating this catalogue, and the Conacyt Project CB-285080 for supporting them.

\textit{Software}: KINEMETRY \citep{Kaviraj_2007}, GILDAS \citep{Pety_2005,GildasTeam_2013}, GALFIT \citep{Peng_2002,Peng_2010}, $^{\rm 3D}$BAROLO \citep{DiTeodoro_2015}.




\bibliography{S0}{}
\bibliographystyle{aasjournal}



\end{CJK*}
\end{document}